\documentclass[conference]{IEEEtran}

\usepackage[switch]{lineno}
\usepackage{xcolor}

\usepackage{graphicx}
\graphicspath{{images}}
\DeclareGraphicsExtensions{.pdf,.png}

\usepackage{amsmath, amssymb}
\usepackage{soul}
\usepackage{algorithmic}
\usepackage{graphicx}
\graphicspath{{./images/}}
\usepackage{textcomp}
\usepackage{xcolor}
\usepackage{import}
\usepackage{tikz}
\usetikzlibrary{arrows, arrows.meta}
\usepackage{hyperref}

\usepackage[font=small]{caption}
\usepackage{subcaption}

\usepackage{tikz,pgfplots}
\pgfplotsset{compat=1.10}
\usepgfplotslibrary{fillbetween}

\newcommand{\fcn}{fpgaConvNet}
\newcommand{\ee}{Early-Exit}
\newcommand{\owl}{ATHEENA}

\newenvironment{customlegend}[1][]{%
    \begingroup
    \csname pgfplots@init@cleared@structures\endcsname
    \pgfplotsset{#1}%
}{%
    \csname pgfplots@createlegend\endcsname
    \endgroup
}%
\def\addlegendimage{\csname pgfplots@addlegendimage\endcsname}
\newcommand{\addlegendimageintext}[1]{%
    \tikz {
        \begin{customlegend}[
            legend entries={\empty},
            legend style={
                draw=none,
                inner sep=0pt,
                column sep=0pt,
                nodes={inner sep=0pt}}]
        \addlegendimage{#1}
        \end{customlegend}
    }%
}
\DeclareRobustCommand\bram{\addlegendimageintext{color=black,mark=x,only marks,style={mark size=4pt}}}
\DeclareRobustCommand\lut{\addlegendimageintext{color=black,mark=square,only marks, style={mark size=3pt}}}
\DeclareRobustCommand\dsp{\addlegendimageintext{color=black,mark=o,only marks, style={mark size=3pt}}}

\begin{document}

\title{ATHEENA: A Toolflow for Hardware Early-Exit Network Automation}

\author{\IEEEauthorblockN{Benjamin Biggs,
Christos-Savvas Bouganis,
George Constantinides
}
\IEEEauthorblockA{Department of Electrical \& Electronic Engineering\\
Imperial College London\\
Email: benjamin.biggs15@imperial.ac.uk,
christos-savvas.bouganis@imperial.ac.uk,
g.constantinides@imperial.ac.uk}
}

\maketitle

\begin{abstract}
The continued need for improvements in accuracy, throughput,
and efficiency of Deep Neural Networks has resulted in a multitude of methods that make the most of custom architectures on
FPGAs. These include the creation of hand-crafted networks and
the use of quantization and pruning to reduce extraneous network
parameters. However, with the potential of static solutions already
well exploited, we propose to shift the focus to using the varying
difficulty of individual data samples to further improve efficiency
and reduce average compute for classification. Input-dependent computation allows
for the network to make runtime decisions to finish a task early if
the result meets a confidence threshold. Early-Exit network architectures have become an increasingly popular way to implement
such behaviour in software. 

We create \textit{A Toolflow for Hardware Early-Exit Network Automation} (ATHEENA), an automated FPGA toolflow
that leverages the probability of samples exiting early from such
networks to scale the resources allocated to different sections of the network. The toolflow uses the data-flow model of fpgaConvNet,
extended to support Early-Exit networks as well as Design Space
Exploration to optimize the generated streaming architecture hardware with the goal of increasing throughput/reducing area while maintaining accuracy.
Experimental results on three different networks demonstrate a throughput increase of $2.00\times$ to $2.78\times$ compared to an optimized baseline network implementation with no early exits. Additionally, the toolflow can achieve a throughput matching the same baseline with as low as $46\%$ of the resources the baseline requires.
\end{abstract}

\section{Introduction}
Convolutional Neural Networks (CNN) have many applications, especially in computer vision and image classification tasks~\cite{CNN_deep_classif}. Traditional CNNs are composed of common operations/layers that can be represented by frameworks like the Open Neural Network Exchange (ONNX)~\cite{onnxSite}.
The continued increase in the width and depth of CNNs has made many of the top performing networks prohibitively large for acceleration on the limited resources of FPGA hardware. There have been a wide variety of static methods~\cite{dnn_approx_survey_wwbwwg} derived to combat the large memory footprints and the compute power required to execute these networks including pruning~\cite{pruning_eg,sparsenet_prune,bayes_bits}, quantization~\cite{BFP_for_effNet, flexor_frac, FINN_single_bit, ternary_nets,lognet}, and knowledge distillation~\cite{kd_eg}. 
These methods require the assumption that the full networks have some level of redundancy that is exploitable across the majority of the data~set. 

A parallel trend to create networks that can better fit on smaller target devices has resulted in more stripped down architectures both hand crafted~\cite{mobnetv2} and generated by Neural Architecture Search (NAS)~\cite{tan2020efficientnet, mcunet2020}. The result is that models are tending to be less redundant, reducing the potential improvement due to methods like pruning.

This is where \textit{input-dependent} computing can take over. The fundamental concept is that a given data sample can be more or less difficult for the network to classify. Practically, this means that data sample A can be accurately classified based on features derived earlier in the CNN, whereas data sample B is more challenging so requires the more refined features of a higher capacity network, resulting in more computation. A number of network architectures make use of this idea to adapt to the computational requirements of individual samples~\cite{branchynet, msdnet_2018_pytorch} at run time. It is possible to reduce the average compute for inference of a multitude of tasks for the relatively low overhead of a calculation to determine the confidence in result. Since the difficult samples can continue through the full network, there is a minimal effect on accuracy and in some cases an accuracy improvement over the baseline networks. 

\begin{figure}[h]
\begin{center}
  \includegraphics[width=0.75\linewidth]{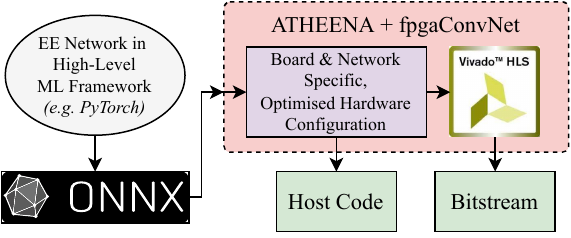}
  \caption{High-level overview of the \owl\ and \fcn\ toolflow.}
  \label{high_level_toolflow}    
\end{center}
\end{figure}

We target throughput-oriented applications that are subject to latency constraints prohibiting device reconfiguration at runtime. We present the following novel contributions for FPGA implementation of such applications that benefit from an input-dependent approach:
\begin{itemize}
    \item A methodology for utilizing probability profiles to select different points on the throughput / area tradeoff curve for different stages of an \ee\ (EE) network.
    \item A set of hardware-friendly components for building early-exit networks on FPGAs, compatible with the open source \fcn\ toolflow.
    \item The \owl\ automated toolflow (Figure~\ref{high_level_toolflow}) for utilizing profiling probabilities to transform their ONNX-based representation into optimized HLS code suitable for implementation using Vivado HLS. 
\end{itemize}

\section{Background}
\label{bg_total}

Section~\ref{bg_id-comp-dnn} outlines existing work on input-dependent deep neural network (DNN) computation. We also summarize the well-studied network we use for our experimental study in Section~\ref{brn_case_study}. 

\subsection{Input-dependent Computation in DNNs}
\label{bg_id-comp-dnn}

The aim of input-dependent computation is to reduce the computational workload of inference whilst maintaining the accuracy of the network. Static compression methods also reduce computational workload but result in accuracy degradation across all inputs. The introduction of input-dependent, dynamic behaviour customises the computational workload per input, allowing the user to explore the trade-off between accuracy and computational workload. This concept is demonstrated in Dynamically Throttleable Neural Network (TNN)~\cite{dyn_thrott_nn}. The network uses a two-stage training approach which first produces a high accuracy network followed by reinforcement learning to train a small DNN module adjacent to the main DNN that has a fine grained control of customizable subsets of the main DNN layers. 
Similarly, Dynamic Deep Neural Network (D$^2$NN)~\cite{d2nn} constructs a network trained with reinforcement learning but has a network topology consisting of `regular' (convolution and fully connected layers) interspersed with `control' nodes which dictate the path a given data sample will take. 

An alternative to the architectural freedom of D$^2$NN are the split computing~\cite{split_compute_early_exit_survey} methods. These are located mid-way on a sliding scale of division of computation between edge device and server: at one extreme, there is fully local computation, where inference is performed on an embedded device, and at the other extreme, data is captured and compressed locally~\cite{split_comp_eg} before being transmitted via a wireless connection to server for inference.
In general, the local computation performs some initial classification (or other ML task) and assesses the confidence of the result. The local compute can then decide whether or not further computation is required and can potentially skip the high latency, round trip of transferring data to power-intensive, server-based compute~\cite{dynamic_edge_eg}. 

\begin{figure}
\begin{center}
  \includegraphics[width=0.95\linewidth]{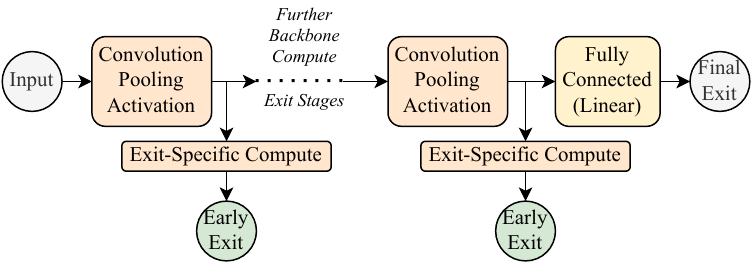}
  \caption{The form of a generic \ee\ network with backbone stages and varying levels of compute between exits.}
  \label{generic_ee}    
\end{center}
\end{figure}

\ee\ networks share structural similarities with split computing and have drawn increasing interest~\cite{ee_survey_samsung} in recent years. The typical architecture of an \ee\ network is set out in the BranchyNet~\cite{branchynet} work and consists of a branching, tree-like structure with a backbone, where the majority of the sample processing is carried out, and exits, which often contain some additional CNN layers and are located at different points along this backbone as illustrated in Figure~\ref{generic_ee}.
Due to the hierarchical nature of CNNs~\cite{msdnet_2018_pytorch,dlbook}, the earlier layers in the backbone will have learnt more general or coarser features. As a result, easy samples can be classified based on these features to an acceptable level of accuracy and more complex samples can be further processed in later stages of the backbone before final classification. 

The benefit of \ee\ methods over traditional quantisation and pruning is that the former make use of the varying difficulty of samples within a data set. Reducing precision or removing extraneous parameters common to all samples beyond a certain point will decrease the accuracy of the network over the most challenging samples first. With the implementation of early exits, inference efficiency can be improved beyond these limits. \ee\ presents a tuneable trade-off between the throughput and accuracy of data samples. The other key benefit of these networks is the improved throughput of batch computation due the average of the reduced latency of early exits and similar latency of later exits. Furthermore, \ee\ has applications in an array of ML tasks including semantic segmentation~\cite{mess} and image classification~\cite{learning_to_stop,triple_wins}. 

The `difficulty' of a given data sample is challenging to quantify so there is a range of metrics used in the literature to judge the confidence of a result at a given exit point. A limitation of the deployment of these networks is the resource/latency cost of these exits which may need to compute exponential or logarithmic operations. The most common method of determining confidence is to measure the information entropy~\cite{branchynet,ee_survey_samsung} of the probability distribution over the available classes. A low entropy value indicates more certainty in the correct result. The other main method to determine confidence is comparing the maximum value in the class probability distribution to a threshold~\cite{msdnet_2018_pytorch}. Both these methods require \texttt{Softmax} computation and the use of logarithmic or exponential functions. 

A key challenge in migrating these software-oriented \ee\ designs to hardware, is to develop an architecture that efficiently executes input-dependent control-flow whilst minimising extraneous computation, area overhead, and maximising the available throughput gains. This is the central challenge we address with \owl\ in this paper, for the case of \ee\ networks.

\subsection{CNN compilers for FPGAs}
\label{bg_cnn-fpga}

Taking standard CNNs from software-based inference to accelerated, FPGA-based inference can be accomplished automatically with custom compilers~\cite{cnn2fpga_survey}. These broadly fit into two categories: single computation engine architectures and data-flow streaming architectures. 
The single engine~\cite{angel_eye} typically consists of a fixed architecture on which the CNN layers are mapped, loaded, and executed in a sequential fashion. The CNN is translated to a list of instructions. This execution can be controlled either by software using CPU or specialised control hardware. The benefit of single engine is that they are amenable to different CNN workloads. However, there is significant benefit, in terms of resource savings and throughput increases, to customising a given architecture to the CNN workload~\cite{overlay_efficiency}. Streaming architectures take this customizability further by making use of the data-flow paradigm to produce deeply pipelined designs with specialized layers for state-of-the-art CNNs~\cite{hpipe_fcnlike}. 

We adopt the streaming architecture method so that we can benefit from CNN-specific customization and implement the input-dependent compute spatially, targeting high throughput. 

\subsection{fpgaConvNet}
\label{bg_fcn_exp}
\fcn~\cite{fpgaconvnet} is an automated CNN-to-FPGA toolflow able to handle common CNN layers and forms a foundation for our toolflow.
The streaming architecture used comes from modelling the CNN as a Synchronous Data-Flow Graph (SDFG)~\cite{sdf}. The nodes are the computation operations such as convolution and the arcs are streams of data. This means that the hardware blocks mapping these computations can have a static schedule and compensate for differing data consumption rates with sufficient buffering determined at compile time. Streaming backpressure is handled by the Vivado HLS~\cite{vivado_hls_ref} streaming interface.

\fcn\ receives a trained CNN model in the ONNX~\cite{onnxSite} representation. This device-agnostic representation is parsed to construct the SDFG and populate an initial hardware mapping with templated layers/modules corresponding to the supported ONNX operations (such as convolution and pooling). 
The tool performs Design Space Exploration (DSE) to optimize the hardware architecture using simulated annealing to select possible incremental transformations to the hardware blocks. This allows an optimized design to be found in a large search space in a relatively short time. Finally, the hardware code is generated in a form suitable for Vivado HLS to compile. The design space exploration makes the most of the flexibility of FPGAs. It makes use of resource models of the hardware blocks as well as the target board resources and the model of the mapped CNN workload. The toolflow is also able to accommodate different application objectives since it is possible to optimize the design for latency or throughput. 

For larger CNNs, folding is used to tune the amount of intermediate computation required to compute a full result. Folding the inputs and intermediate feature maps multiplexes sections of the design in time to reduce resource consumption. This is accomplished at two scales: coarse-grain folding at the input and output of layers and fine-grain folding of the sliding windows within the convolution module. 

\fcn\ was chosen as a basis for our work because of the existing hardware templates for fundamental CNN layers and the tooling infrastructure that provides DSE to generate a network layer configuration for hardware implementation. \fcn\ is open source and demonstrably extensible~\cite{samo_fpgaconvnet_ext, streamsvd_fpgaconvnet_ext, power-aware_fpgaconvnet_ext} in a range of contexts. In this work, we transform \fcn\ into \owl\ by extending the scope of supported CNN architectures as well as creating new hardware templates to efficiently support \ee\ networks. As with existing comparable toolflows~\cite{hpipe_fcnlike,hls4ml_bin_nn, finn} (at time of writing), \fcn\ does not support the coarse granularity of input-dependent computation of \ee\ networks. We expand the data-flow model of \fcn\ to include input-dependent computation as pipelined control flow. This allows \owl\ to generate CNN-hardware mappings that benefit from varying data sample difficulty. The method of extension we develop for \fcn\ is orthogonal to extreme quantization~\cite{finn} and exploitation of sparsity~\cite{hpipe_fcnlike} detailed in other works. Furthermore, we improve overall compilation times by partitioning the design prior to HLS compilation and automatically stitching generated components prior to synthesis and implementation.

\subsection{Dynamic Machine Learning on FPGAs}
\label{bg_related}
CascadeCNN~\cite{cascadecnn,cascadecnn_upgrade} implements a specialized form of dynamic computation where a runtime decision is made to switch between low and high-precision quantized versions of the same network. The components of the architecture include both a low and high precision implementation of the network on FPGA fabric and the confidence evaluation unit running on a CPU. All data examples are first fed through the low precision network before the computation of a confidence estimate on the results to determine whether the results meet the user-defined accuracy threshold. Any samples with low confidence are fed through the high precision network. \owl\ is similar in that all data sample processing compute is confined to the FPGA fabric avoiding prohibitive reconfiguration costs~\cite{cascadecnn_upgrade}. We also choose to include the confidence evaluation on chip for \owl, meaning data does not need to make the round trip to and from off chip memory to perform the confidence decision. The key difference for \owl\ is the streaming architecture which can achieve higher throughput thanks to deeply pipelined layers.

DynExit~\cite{DynExit} uses a hand crafted, single engine architecture to implement a ResNet with classifiers attached along a pre-trained backbone. The network architecture consists of pipelined convolution and linear (fully connected) execution blocks attached to a `branch'. The `branch' consists of an exponential and a natural logarithm module to compute the confidence of the classification based on a rearranged version of the cross entropy loss function. Similarly, \owl\ uses a dedicated block to determine the sample confidence but DynExit lacks customizable layer configurations which limits the accelerator design's ability to fully utilise the FPGAs resources for different networks.

Adaptive Hierarchical CNN (AHCNN)~\cite{AHCNN_farhadi_fpga} notes the benefits of being able to utilise the shallow features for easy data sample classification and deeper features of a more accurate network for difficult classifications. To this end, partial reconfiguration is used to swap in and out shallow and deeper sections of a ResNet18 CNN. Large batch sizes are used to amortise the latency penalty for reconfiguration. This allows the design to benefit from full utilization of the resources for each stage of network computation but the latency penalty is prohibitive for low latency applications. They also compute the confidence decision based on the maximum value of the \texttt{Softmax}, as with \owl, but couple this with the profiled probability of the occurrence of a given class. This equates to a confidence threshold that is class-dependent.

The heterogeneous architecture proposed in~\cite{ee_hw_sw_codesign} makes use of the FPGA fabric to accelerate the highly parallel CNN computation through the use of multiple processing elements in a systolic array architecture. The onboard CPU computes the \texttt{Softmax} and entropy of the intermediate classification results. As with DynExit, this is another hand-crafted architecture that supports the main convolution kernel sizes but currently has limited support for more recent networks.

Interest is growing in the use of input-dependent computation, however, none of these works provide a fully automated toolflow for mapping \ee\ CNNs to FPGAs. It is this problem we tackle in this paper.


\section{Methodology}
The fundamental issue of implementing \ee\ networks on FPGAs is determining a hardware mapping that balances the control and data-flow throughput requirements whilst minimising latency overhead. A naive implementation would have all stages of the network optimized for the highest possible throughput. 
However, in the presence of any resource constraints this is clearly a sub-optimal strategy: having all stages targeting the same fixed throughput will lead to some stages being under-resourced bottlenecks and others being starved of data samples.
Hence in this work we target the following problem: given an FPGA platform with certain computational and memory resources, what is the best way to allocate those resources to maximise throughput for a given expected distribution of samples of varying difficulty.
We first define our methodology for determining this optimized resource allocation for \ee\ networks, and then explain the detail of the toolflow and template hardware designs we introduce to \fcn\ for \owl. The code will be open-sourced\footnote{Repository DOI: \texttt{\url{https://doi.org/10.5281/zenodo.7809222}}}. 

\subsection{Scaling Resource Allocation according to Exit Probability}
\label{meth_area_prob}

\begin{figure}
\begin{center}
  \includegraphics[width=0.9\linewidth]{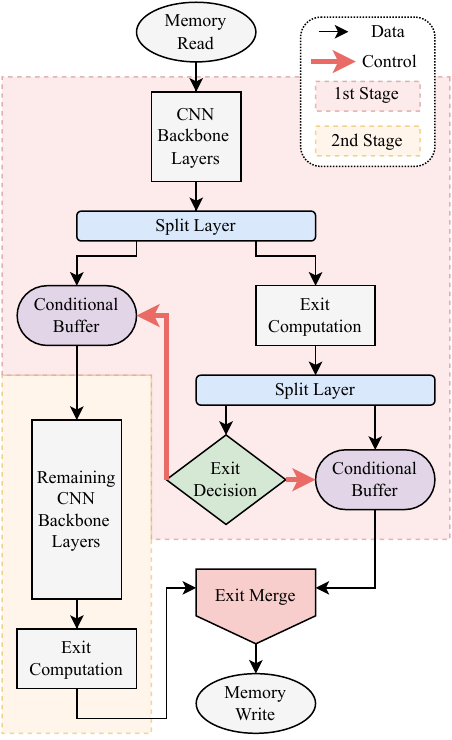}
  \caption{High-level diagram of the proposed control-flow hardware attached to the CNN layers of \fcn\ for a two stage network. Black arrows represent normal data-flow plus Sample ID tags and red arrows represent control signals.}
  \label{fcn_ctrl_hw}    
\end{center}
\end{figure}

\begin{figure}
\begin{center}
  \includegraphics[width=\linewidth]{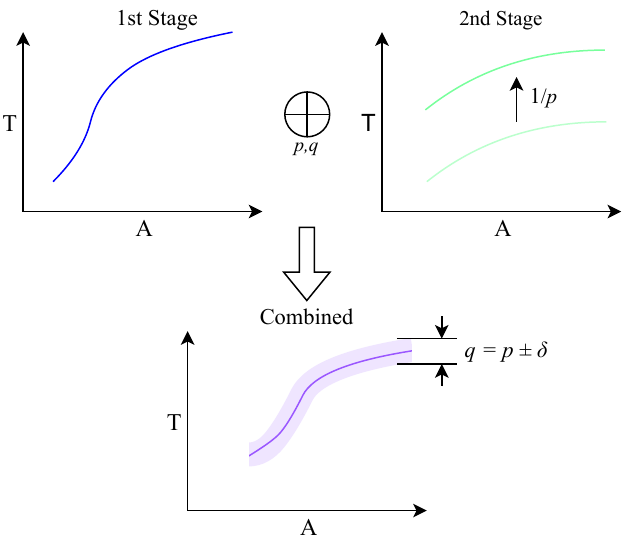}
  \caption{Visual representation of the scaling and combination of TAP curves for network stages. Our process combines two TAP graphs with respect to an expected probability $p$. This $p$ corresponds to the percentage of samples that require processing by second stage. $q$ denotes throughput deviations from this design-time probability that may be encountered at runtime.}
  \label{ta_pareto_meth}    
\end{center}
\end{figure}

\ee\ networks can be divided into sections according to the stages of compute between each exit. For ease of presentation, we explain the area apportioning process with reference to a two-stage network, however it is trivial to extend the presentation to multi-stage networks. We illustrate the methodology with the generic, two-stage network in Figure~\ref{fcn_ctrl_hw} before applying it to BranchyNet in Section~\ref{brn_case_study}. The network is first separated into two stages at the layer level. This means dividing the network into the first stage, containing all the parts of the backbone and the \ee\ layers that are need to operate at the higher data rate, and the second stage containing the remaining parts of the backbone and final exit. This second stage is only required to operate at a lower data rate because not all input data samples pass through this hardware. As a result, network classification decisions may also return out-of-order. The key challenge is to automate the design of a hardware architecture capable of efficiently supporting these multiple data rates. To this end, we can use existing FPGA design tools capable of folding (in our case \fcn) to generate separate, optimized Throughput-Area Pareto (TAP) functions for both stages which we then merge to generate a combined TAP function as visualized in Figure~\ref{ta_pareto_meth}.

We define a TAP function as a function that is (non-strictly) monotonically increasing in each of its arguments. The principle is that this function captures the maximum achievable throughput possible by separately optimizing a section of the network for a given fraction of total resources. 
Let $\mathbb{N}$ denote the set of natural numbers and $\mathbb{Q}$ denote the set of rational numbers. An example of a TAP function might then be,
$f : \mathbb{N}^4 \to \mathbb{Q}$, capturing  the optimal throughput possible with a constrained number of BlockRAMs, DSPs, FFs, and LUTs, as represented by the four arguments to the function. A function like this is automatically generated by providing the \fcn\ optimizer limited fractions of the board resource constraints. The results for each set of constraints are collated for input to the \owl\ optimizer. The \textit{1st Stage} and \textit{2nd Stage} graphs of Figure~\ref{ta_pareto_meth} show a sketch of some TAP functions for the first and second stage of a two-stage network. 

Our objective is to combine these two TAPs into a single TAP for the \ee\ network. We may think about the problem in this way: since the second stage is only expected to be used by each input sample with some probability $ p \in (0,1] $, it follows that, given suitable buffering between stages, it is possible to extract a higher throughput than the nominal throughput of the design, by a factor $1/p$. 

The overall design will be limited by the throughput of the first stage or the second stage scaled by $1/p$, whichever is lower. However, in any practical setting, the probability for an input sample to need further processing by the second stage will differ by some degree from the (profile-based) probability for which the hardware was designed. We therefore denote the design-time probability estimate as $p$ and the actually encountered probability as $q$. Putting these components together, we are able to define an ideal combination of the two TAPs, parameterised by $p$ and $q$, expressed formally below, and illustrated in Figure~\ref{TAP_comb_bound} where $\oplus$ denotes the so-defined TAP combination operator.

\begin{equation}
\label{TAP_comb_bound}
\begin{aligned}
    f \underset{p,q}{\oplus} g: x \mapsto \min(f(x_1),g(x_2)/q) \\
    \mathrm{ where } \; (x_1,x_2) = \underset{\underset{\textrm{ s.t. } x = x_1 + x_2}{x_1,x_2} }{\arg\max}  \min( f(x_1), g(x_2)/p)
\end{aligned}
\end{equation}

Intuitively, what this equation tells us is that for a given total resource budget $x$, we should apportion a resource allocation $x_1$ for the first stage and $x_2$ for the second stage maximising the throughput of the limiting stage, taking into account our design-time estimate of probability $p$. At runtime, the throughput demand on the first stage will be as expected, but may vary somewhat from the design-time expectation on the second stage. This is all illustrated in the lower plot of Figure~\ref{ta_pareto_meth}. The design points represented by the TAP function for the first and second stages are discrete. This means that it is unlikely for the tool to be able to perfectly match the predicted throughput values. In the case that the second stage is the limiting factor, a reduction in the use of this stage, corresponding to $q < p$, will result in an increase in throughput. For $q > p$, the throughput will be reduced due to the reliance on the limiting stage. These situations are represented by the shaded region. The solid purple line corresponds to $q = p$ , in which the probability of hard samples matches that of the profiled, design-time probability.

The following section explains how we obtain our probability profiles and use them to inform the combination of TAP functions in the manner previously described. \owl\ builds on \fcn\ and automates this process for the user. We demonstrate this process in practice in Section~\ref{sec_eval}. 

\subsection{Toolflow Extensions \& Automation}
\label{meth_auto_tool}

\begin{figure}
\begin{center}
  \includegraphics[width=0.95\linewidth]{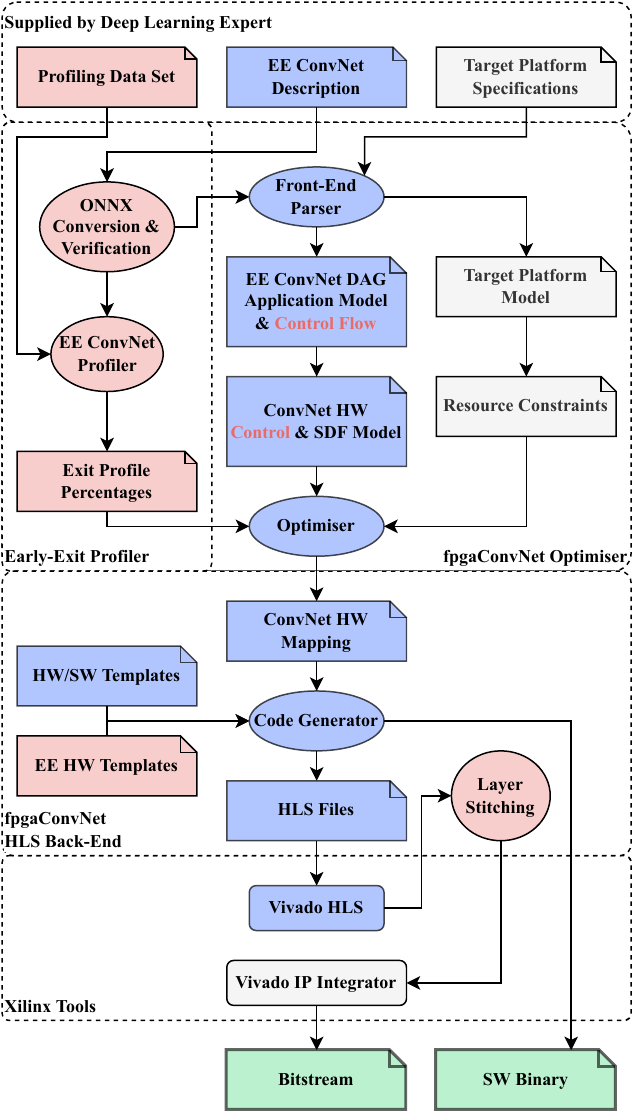}
  \caption{A visual representation of the \owl\ toolflow. We have built upon the baseline, open source \fcn\ toolflow. New elements are in red and modified elements are in blue.}
  \label{fcn_toolflow_changes}    
\end{center}
\vspace*{-4mm}
\end{figure}

We build the \owl\ toolflow by extending the open-source \fcn\ as illustrated in Figure~\ref{fcn_toolflow_changes}. The fundamental difference between the flows is that the original \fcn\ constructed a data-flow graph whereas we require a \textit{control} and data-flow graph (CDFG) to represent the flow of data through layers as well as the confidence decisions at the end of the early exits. Modifications to the parser and optimizer are made to support the different ONNX operations and encompass the control-flow for hardware translation. Several new hardware component templates (detailed in Section~\ref{meth_ee_layers}) are designed for the FPGA implementation of control-flow and to support the confidence calculation. 

\subsubsection{\ee\ Profiler}
We introduce the \ee\ profiler which takes a \textit{profiling data set} and the high-level \textit{\ee\ ConvNet description} and apportions the set so that multiple distinct tests can be run which will have a similar probability of hard samples on average but variation individually. Batched inference is performed over the sets followed by collection of the exit probabilities, exit accuracy, and cumulative accuracy. The average probability of hard samples is fed into the optimizer as $p$, along with the multi-stage CDFG hardware model. The \owl\ optimizer then creates an optimized mapping for the different stages of the network, scaling resource constraints according to $1/p$. 

\subsubsection{HLS Limitations \& Parallel Compilation}

In order to allow for HLS-based compilation of large networks, we  automatically split the network into the individual layers, generating top-level HLS files for each. This results in multiple smaller designs for HLS that can be compiled independently.  
The layers are then automatically stitched together at the board design stage in Vivado IP Integrator in conjunction with the supporting processor and memory interfaces. A DMA controller is introduced with input and output FIFOs to manage the transfer of data between the host and the FPGA.
Since the \ee\ board design now consists of multiple HLS cores, they each need a start signal from the host CPU. 
These signals are automatically added into the host code, in addition to the DMA read and write transfers set to batch size specified by the user. By the end of the compilation, the user has a bitstream and complementary host code to perform batch inference on the board.

\subsubsection{ONNX Conversion}
We make some minor modifications to the benchmark source code so that it can be converted into a compatible ONNX form using our \textit{\ee\ profiler}. As the original networks were typically run in software, PyTorch (version 1.8.1) handles the scheduling of the network graph execution for CPU and GPU inference. ONNX generation from a PyTorch description of simple network is trivial but the inclusion of conditional operations requires the PyTorch description to explicitly prevent the operations from being removed by software optimization passes. These network control-flow decisions need to be captured and translated to FPGA-based hardware blocks. We use the PyTorch scripting methods as an intermediate stage~\cite{pytorchscript}. This converts the network into a PyTorch-specific intermediate representation capable of supporting conditional statements. PyTorch-based ONNX methods then convert the intermediate representation to the final ONNX form. An inference test is performed with the ONNX form and the results compared to the original to verify the conditional functionality matches that of the PyTorch implementation. 

\subsection{\ee\ Network Layers: Hardware Templates}
\label{meth_ee_layers}

We extended the ONNX parser of \fcn\ to support the operations required by the \ee\ CNNs. These operations include:

\begin{itemize}
    \item \texttt{Softmax}
    \item Reduction (\texttt{ReduceMax})
    \item Numerical comparison (\texttt{Greater than})
    \item \texttt{If} conditional
\end{itemize}

\begin{figure*}
\begin{center}
  \includegraphics[width=0.9\linewidth]{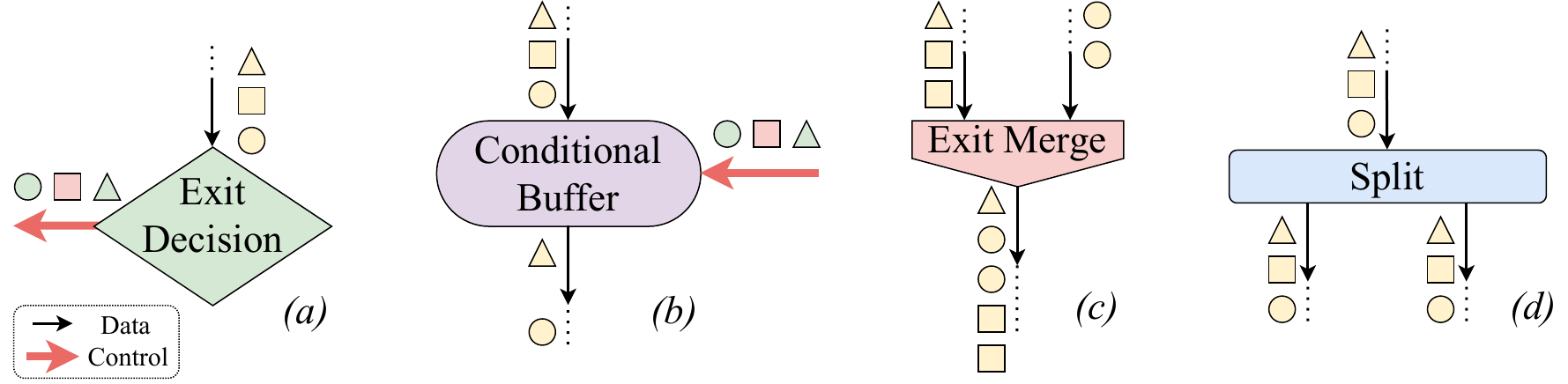}
  \caption{Newly added \ee\ Hardware layer templates. The shapes represent a given Sample ID. For the Conditional Buffer and Exit Decision, we have indicated whether the Sample ID is kept (green) or dropped (red). The associated Sample data is either passed through or dropped. We show that the Exit Merge layer keeps all data for a given Sample ID sequential, opting to stall an exit instead of interleaving.
  }
  \label{ee_hw_layers}    
\end{center}
\end{figure*}

Figure~\ref{ee_hw_layers} illustrates the new \ee\ layer templates and Figure~\ref{fcn_ctrl_hw} provides a high-level view of the proposed placement of these layers within the pre-existing CNN data-flow operations in \fcn. 

These operations are the foundation of the conditional aspect of the network so are merged into one hardware layer comprising of distinct modules. The remaining layers added do not correspond to ONNX operations but support the control flow extensions of the SDF paradigm. The newly added layers match the fixed-point representation with the exception of the Exit Decision layer. This instead uses single-precision floating-point as this preserves the numerical behaviour of the exponential function at its core.

\subsubsection{Exit (Softmax) Decision Layer}
\label{sftmx_resrc_sec}
An early exit will occur if Condition~(\ref{exit_cond}) holds for some threshold $C_{thr}$ determined after training prior to exit profiling, where the standard $\textrm{Softmax}: \mathbb{R}^C \to \mathbb{R}^C$ function corresponds to~(\ref{softmax}), used to transform the vector of class activations into a probability distribution, where exponentiation of vectors is interpreted component-wise~\cite{dlbook}.

\begin{equation}
\label{exit_cond}
    \max_{i \in \{1\ldots C\}}\left[\textrm{Softmax}(x)\right]_i > C_{thr}
\end{equation}

\begin{equation}
\label{softmax}
    \textrm{Softmax}(x) = \frac{\exp(x)}{\sum^{C}_{j=1} \exp(x_j)}
\end{equation}

\noindent $C$ is the number of classes, $x_{i}$ is the network output for the $i$th class, and $C_{thr}$ is the confidence threshold. 

\begin{figure}
\begin{center}
  \includegraphics[width=0.75\linewidth]{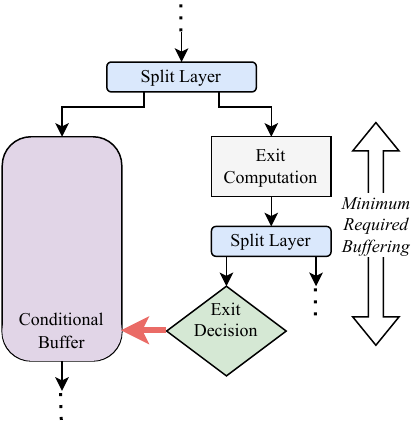}
  \caption{The latency of the additional exit computation and exit decision layers is used to determine the minimum amount of buffering required by the conditional buffer to prevent deadlock in the design. 
  }
  \label{min_buff}    
\end{center}
\vspace*{-3mm}
\end{figure}

This layer, along with the exit-specific computation required for the \ee\ classifier dictate the 
minimum size of buffering required between the intermediate result and the conditional buffer layer to prevent a pipeline stall (Figure~\ref{min_buff}). On a small FPGA this has the potential to be a prohibitively large amount of memory required so the need to reduce the latency of the operation is key to the implementation of \ee\ networks. 

The operation can be rearranged to remove the division as in (\ref{exit_cond_divless}) to decrease the resource overhead and latency of the decision component. The use of floating-point for this layer means that addition and comparison operations incur a significant latency penalty given a target frequency. To reduce this penalty, we implement adder and comparison trees to compute results in parallel, minimising the latency of this layer.

\begin{equation}
\label{exit_cond_divless}
    \max_{i \in \{1\ldots C\}}\exp(x_i) > C_{thr} \cdot \sum^{C}_{j=1} \exp(x_j)
\end{equation}

\subsubsection{Conditional Buffer Layer}
Two challenges arise from the addition of control flow from conditional buffers. 
The first challenge is that, after the first stage of the network, data samples will go down the path to the early exit or the path to the second stage. Let's assume $N$ data samples, $K$ will exit early while $N-K$ will continue through the second stage. The existing \fcn\ hardware layers in the second stage expect $N$ data samples and the implementation of the pipeline will only terminate on the final $N$th data sample. 
To avoid deadlock, we flush the pipeline with an unused sample ID and corresponding data.

The second challenge is that there is a significant amount of computation between the point at which the network first has to buffer data and the point at which control signals are produced. We temporarily buffer the unfinished data sample while the confidence metric is evaluated as well as the fully processed sample result at the early exit classification stage. If the sample can exit early, the buffer drops the intermediate data and the classification result of the early exit is transferred to memory via the exit merge component. Due to the size of the intermediate feature map buffer, it is important to reduce the impact of both reading in and writing out of the buffer (effectively doubling the latency of the layer for every sample prior to the control signals). To drop an unused feature map we invalidate the addresses of the stored feature map in a single cycle.
If the sample cannot exit, the buffer passes the intermediate data through to the next stage of the backbone. The fundamental operation of the conditional buffer is to filter the easy samples from the hard samples. The buffer prevents the later stages of the backbone from unnecessary computation, which has the effect of increasing throughput because of the lower expected data rate for the second stage. This component will be included in the open-source repository. 

\subsubsection{Split Layer}
The split layer is used to duplicate the result of layers at the branching points in the \ee\ network. This splits the data-flow stream to allow a copy of the data to continue down the backbone in parallel to the early exit layers. 

\subsubsection{Exit Merge Layer}
Inference is run on a batch of data samples where each data sample consists of a fixed number of pixels. In line with other static streaming architectures, the original \fcn\ has no built-in distinction between data samples in the pipelined hardware. Each component will continue to operate on newly provided pixels and the user is responsible for interpreting the results based on their location in memory. However, given that in an \ee\ network, data samples within a batch are able to complete out of order, there needs to be an internal representation of each data sample's position within the batch. A \textit{Sample ID} is assigned to each data sample and is passed through the hardware with each pixel. At the conditional buffers, the IDs are compared to determine whether or not to drop the data points with a given ID. When a sample exits the network, the exit selection layer coherently merges the exit streams into one memory writing component. This may result in the stalling of one network path while another is allowed to pass through. 

\section{Experimental Results}
\label{sec_eval}
\begin{figure}
\begin{center}
  \includegraphics[width=0.95\linewidth]{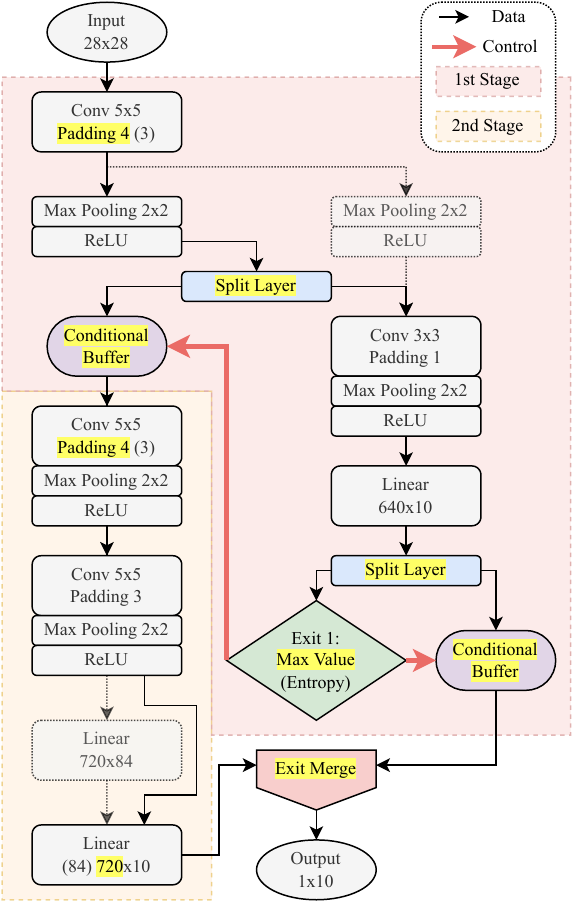}
  \caption{Slightly modified version of B-LeNet~\cite{branchynet} for \fcn. Changes are highlighted in yellow with original values in brackets. Faded layers have been removed from the architecture. Hardware-only layers (detailed in Section~\ref{meth_ee_layers}) have been added.}
  \label{b_lenet_fcn_comb}    
\end{center}
\vspace*{-3mm}
\end{figure}

The following section details a case study of using the \owl\ toolflow to automatically implement the previously-proposed Branchy-LeNet network~\cite{branchynet} directly from PyTorch code. This starts with some minor modifications to the source architecture shown in Figure~\ref{b_lenet_fcn_comb}. The network was trained and tested in the manner outlined in the original paper. The alterations resulted in a negligible change in accuracy and a similar \ee\ probability for a comparable $C_{thr}$ value in software. These modifications reduce wasted compute on the board and improve compatibility with standard \fcn\ convolution layers. The \ee\ profiler is used to generate the ONNX representation and \ee\ probabilities for the extended \fcn\ optimizer. This in turn creates a hardware mapping for the \ee\ and \fcn\ layers for the chosen board. The tool then converts the mapping into HLS code and compiles the layers in parallel before stitching them together and generating a bitstream with associated host code. 

\def \baselineGraphColour {red}
\definecolor{combGraphColour}{RGB}{154,87,255}

\begin{figure*}
\begin{subfigure}[b]{0.49\textwidth}
    \begin{tikzpicture}
        \begin{axis}[
            xlabel={Predicted Fraction of Maximum Limiting Resource},
            ylabel={Predicted Throughput [Samples/s]},
            xmin=0.15, xmax=1.00,
            ymin=0, ymax=100000,
            xtick={0.2,0.3,0.4,0.5,0.6,0.7,0.8,0.9,1.0},
            ytick={0,10000,20000,30000,40000,50000,60000,70000,80000,90000,1000000},
            legend pos=north west,
            ymajorgrids=true,
            xmajorgrids=true,
        ]
            \addplot[const plot,color=\baselineGraphColour,thick] 
                table[col sep=comma]
                {data_hw/baseline.csv};

            \addplot[color=\baselineGraphColour,mark=square,only marks,]
                table[col sep=comma]{data_hw/baseline_LUT.csv};
            \addplot[color=\baselineGraphColour,mark=x,only marks,]
                table[col sep=comma]{data_hw/baseline_BRAM.csv};

            \addplot[const plot,color=combGraphColour,thick]
                table[col sep=comma]
                {data_hw/Opt_25_curve.csv};
                
            \addplot[color=combGraphColour,mark=x,only marks,]
                table[col sep=comma]{data_hw/Opt_25_BRAM.csv};

            \addplot[color=combGraphColour,mark=square,only marks,]
                table[col sep=comma]{data_hw/Opt_25_LUT.csv};

            \addplot+[const plot,name path=20, color=combGraphColour,no markers,dashed]
                table[col sep=comma]
                {data_hw/Opt_25_bound_20.csv};
            \addplot+[const plot,name path=30, color=combGraphColour,no markers,dashed]
                table[col sep=comma]
                {data_hw/Opt_25_bound_30.csv};
            \addplot[combGraphColour!20] fill between[of=20 and 30];
        \end{axis}
    \end{tikzpicture}
    \begin{center}
        \caption{Throughput Area results generated by the simulated annealing optimizer stage of the toolflow at a $p$ of 25\% and $q = p \pm 5\%$.}
        \label{area_thruput_optim_board_l}
    \end{center}
\end{subfigure}
\hskip 0.01\linewidth
\begin{subfigure}[b]{0.49\textwidth}
    \begin{tikzpicture}
        \begin{axis}[
            xlabel={Fraction of Maximum Limiting Resource},
            ylabel={Throughput [Samples/s]},
            xmin=0.15, xmax=1.00,
            ymin=0, ymax=100000,
            xtick={0.2,0.3,0.4,0.5,0.6,0.7,0.8,0.9,1.0},
            ytick={0,10000,20000,30000,40000,50000,60000,70000,80000, 90000, 1000000},
            legend pos=north west,
            ymajorgrids=true,
            xmajorgrids=true,
        ]
            \addplot[const plot,color=\baselineGraphColour,thick] 
                table[col sep=comma]
                {data_hw/lenet_se_baseline_HW.csv};
            \addplot[color=\baselineGraphColour,mark=o,only marks,]
                table[col sep=comma]{data_hw/lenet_se_baseline_HW_DSPS.csv};
            \addplot[color=\baselineGraphColour,mark=square,only marks,]
                table[col sep=comma]{data_hw/lenet_se_baseline_HW_LUTS.csv};
            
            \addplot[const plot,color=combGraphColour,thick]
                table[col sep=comma]
                {data_comb/comb_HW.csv};
                
            \addplot[color=combGraphColour,mark=o,only marks,]
                table[col sep=comma]{data_comb/comb_DSP_HW.csv};
            LUT    
            \addplot[color=combGraphColour,mark=square,only marks,]
                table[col sep=comma]{data_comb/comb_LUT_HW.csv};
                
            \addplot+[const plot,name path=20, color=combGraphColour,no markers,dashed] 
                table[col sep=comma]
                {data_comb/comb_upper_HW.csv};
            plot the lowerline
            \addplot+[const plot,name path=30, color=combGraphColour,no markers,dashed] 
                table[col sep=comma]
                {data_comb/comb_lower_HW.csv};
            \addplot[combGraphColour!20] fill between[of=20 and 30];

            \node[color=combGraphColour] (pnt_a) at (axis cs:0.285,23000){A1};
            \node[color=combGraphColour] (pnt_b) at (axis cs:0.551,50000){A2};
            \node[color=combGraphColour] (pnt_c) at (axis cs:0.86,96500){A3};
            \node[color=\baselineGraphColour] (pnt_d) at (axis cs:0.37,9000){B1};
            \node[color=\baselineGraphColour] (pnt_e) at (axis cs:0.545,16800){B2};
            \node[color=\baselineGraphColour] (pnt_f) at (axis cs:0.97,38500){B3};

            \draw[->] (axis cs:0.98,44800)--(axis cs:0.83,92800);
            \node[align=left] (thru_diff) at (axis cs:0.93,85000){Thru:\\$2.17\times$};

            \draw[<-] (axis cs:0.48,43400)--(axis cs:0.973,43400);
            \node[align=left] (rsc_diff) at (axis cs:0.75,50000){Rsc:\\$0.46\times$};
        
        \end{axis}
    \end{tikzpicture}
    \begin{center}
        \caption{Throughput Area results obtained using ZC706 board. At a $p$ of 25\%. Tested using data sets consisting of $q = 30\%, 25\%, 20\%$.}
        \label{area_thruput_optim_board_r}
    \end{center}
    
\end{subfigure}

    \caption{
    The red line represents the corresponding baseline results for \fcn\ and the limiting resources for each point are denoted by the following: \bram   (BRAM), \lut   (LUT), \dsp   (DSP).
    Designs A1, A2, A3, and B1, B2, B3 have detailed resource usage in Table~\ref{b-lenet-rsc}.
    }  
    \label{area_thruput_optim_board}
\end{figure*}
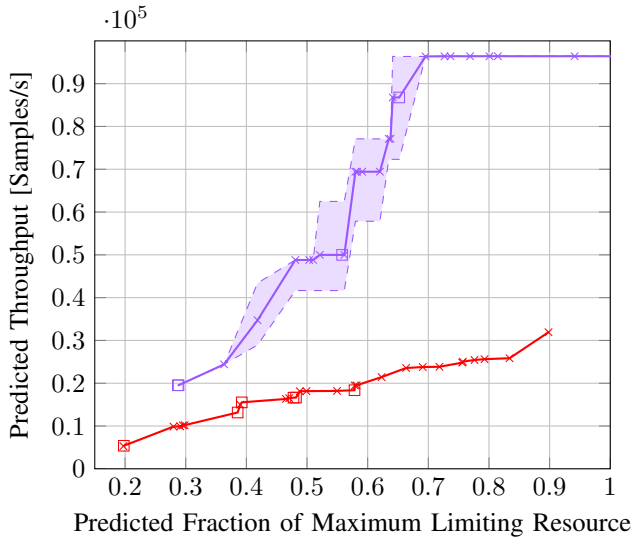
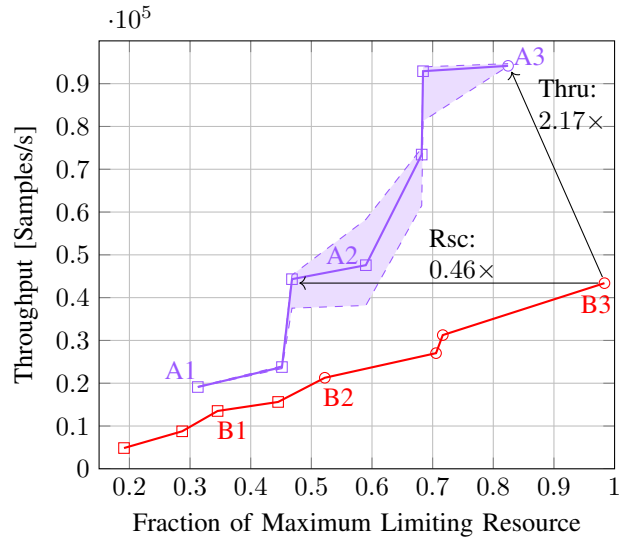

\subsection{BranchyNet: Hardware Experimental Study}
\label{brn_case_study}

The experimental setup follows that of \fcn\ to make comparison direct and expedite the gathering of results. The target device is the Xilinx ZC706 board~\cite{zc706} with the Zynq 7045 System on Chip (SoC). The resources available are 218600 LUTs, 437200 FFs, 900 DSPs, and 1090 18K BRAMs. Vivado HLS 2019.1 was used for compatibility with \fcn. Each design is conservatively clocked at 125MHz (limited by HLS and board). We compare our hardware \ee\ implementation to a corresponding single-stage network baseline. This single-stage (backbone) consists of the network layers from the start of the \ee\ network through to the end of the second stage. For BranchyNet, this means three Convolution, Pooling and ReLU layers followed by a Linear (Fully Connected) layer. 

Both the \owl\ optimizer and baseline optimizer are provided the board resources constrained at different percentages in order to generate a Throughput-Area Pareto curve. Due to the random aspect of the simulated annealing within both optimizers, they are run ten times and the best points are chosen to form the curve. Additionally, a range of data points with constrained resources allows us to infer throughput gains/resource savings on boards with lower available resources. Once the optimizers generate the hardware mapping, we collate the predicted results shown in Figure~\ref{area_thruput_optim_board_l} and then pass through the best performing subset of these results to the HLS backend in order to generate the board design. Figure~\ref{area_thruput_optim_board_r} shows the resulting Throughput-Area Pareto curve after synthesis, implementation and place-and-route. The resource usage is recorded and the board loaded with each bitstream. The automatically generated host code for the board loads a batch of 1024 samples onto off-chip memory and enables the DMA transfers to and from the design. We measure the total time taken from the start of the DMA transfer until the DMA block registers as being idle and use this to calculate the throughput of the network. The same DMA controller is present for baseline and \ee\ implementations so the impact on resources is consistent for a fair comparison.

The baseline results are represented by the red lines in Figure~\ref{area_thruput_optim_board}. While the \fcn\ model is not accurate on a point by point basis, the trend is consistent for the predicted and board implementations. 
To allow for a fair comparison, we included an implemented point predicted by the optimizer to consume greater than the boards resources, in practice the design point B3 consumed $98\%$ of the DSPs. For higher resource allowances, we can see that the designs tend towards being limited by DSPs, this is due to the baseline optimizer selecting an increased level of parallelism for the Convolution and Linear layers and is common to both the baseline and ATHEENA board implementations.

The solid purple line shows the case when the $q$ (percentage of easy samples in the test set)  matches that of $p$ (the profiling data set percentage), in this case a $q = p = 25\%$. The dashed lines represent a range of points taken on the Throughput-Area space with a differing percentage of hard samples.
The predicted results are calculated under ideal conditions: assuming a regular sequence of easy and hard samples and sufficient buffering such that second network stage is able to achieve the estimated throughput. 

For the implemented board results, we sample a batch from the test data set for the task. The sampled test set has a split of easy and hard samples proportioned according to the required test probabilities but distributed randomly within the batch of 1024 samples. The lower dashed line represents a $q = 30\%$ and demonstrates a partially reduced throughput for some of the data points. 
The upper dashed line represents a $q = 20\%$ and shows that some sub-optimal profiling situations can still result in throughput increase thanks to the significant latency reduction of data samples exiting at the high-throughput first stage. 
The results demonstrate the potential for greatly improved throughput at the risk of slightly decreased throughput caused by differences between test and profiling exit probabilities, and variation in exit distribution within a batch. Additionally, the predicted Throughput-Area curve is a good approximation of the measured implementation results. We can see the \owl\ model is slightly optimistic in terms of achievable throughput. This is likely due to sub-optimal resource models for the new components and the variability of HLS compilation. Finally, the gap between lower dashed line and the baseline is indicative of the robustness of the approach to the difference between $p$ and $q$ for a real world application, assuming sufficiently sized buffers. 

\begin{table}
\centering
\caption{Tabulated resource comparison for implemented Baseline Vs ATHEENA on ZC706 board. *Maximum limiting resource.}
\resizebox{0.5\textwidth}{!}{%
\begin{tabular}{c|c|c|c|c|c|c}
\multicolumn{1}{l|}{} & \textbf{LUT} & \textbf{FF} & \textbf{DSP} & \textbf{BRAM} & \textbf{\begin{tabular}[c]{@{}c@{}}Limiting \\ Resource (\%)\end{tabular}} & \textbf{\begin{tabular}[c]{@{}c@{}}Throughput\\ (Samples/s)\end{tabular}} \\ \hline \hline
B1 & 75513* & 61361 & 295 & 55 & 35 & 13513 \\ \hline
A1 & 68383* & 63170 & 169 & 206 & 31 & 19434 \\ \hline
B2 & 105451 & 84761 & 470* & 89 & 52 & 21276 \\ \hline
A2 & 128940* & 117138 & 407 & 239 & 59 & 47583 \\ \hline
B3 & 138194 & 120764 & 885* & 170 & 98 & 43384 \\ \hline
A3 & 163637 & 142913 & 742* & 265 & 82 & 94170
\end{tabular}%
}
\label{b-lenet-rsc}
\end{table}


\begin{table}
\centering
\caption{Resource overhead of the Early-Exit for labelled designs compared to the network backbone. This includes the resource usage attributed to the \textit{additional} Early-Exit computation and buffering required for operation of B-LeNet. 
The proportion of this overhead is detailed as a percentage of the total design.
}
\resizebox{0.45\textwidth}{!}{%
\begin{tabular}{l|c|c|c|c|c|c|c|c}
 & \textbf{LUT} & \textbf{\%} & \textbf{FF} & \textbf{\%} & \textbf{DSP} & \textbf{\%} & \textbf{BRAM} & \textbf{\%} \\ \hline \hline
A1 & 13912 & 20 & 13595 & 22 & 34 & 20 & 114 & 55 \\ \hline
A2 & 37766 & 29 & 35941 & 31 & 122 & 30 & 146 & 61 \\ \hline
A3 & 33166 & 20 & 30974 & 22 & 112 & 15 & 186 & 70
\end{tabular}%
}
\label{rsc-overhead-table}
\vspace*{-3mm}
\end{table}

The maximum measured \owl\ throughput is $2.17\times$ the the maximum measured baseline throughput. \owl\ achieves this throughput using 16\% fewer DSPs (the limiting resource) when the resources are apportioned between the first and second stage according to the profiled probabilities. Alternatively, \owl\ can achieve the same throughput as the maximum baseline using $46\%$ of the design's limiting resource.

The design points in Table~\ref{b-lenet-rsc} show that the achieved throughput increase comes at the cost of an increase in BRAM. The designs A1, A2, and A3 (labelled in Figure~\ref{area_thruput_optim_board_r}) require this as part of the conditional buffers to store enough of intermediate feature map samples to prevent deadlock, as there is a delay from the point the buffered data and the related control signal are produced. 
Furthermore, additional BRAM is added to increase robustness to variation in the hard samples' exit probability. This results in the resource overhead from implementing the additional classifier layers, comparison layer, and conditional buffering layers being dominated by BRAM usage, as detailed in Table~\ref{rsc-overhead-table} with design A3 having $70\%$ of the total BRAM usage within the early exit overhead.

The predicted throughput of each stage of the \ee\ network is calculated separately. The resulting design points for each stage form a discrete Pareto front. When the optimizer is selecting from these two stages and scaling the throughput of the second stage, there will often be a discrepancy between the predicted throughputs of the stages. If the resulting combined design point over-provisions the second stage then the design will be more robust to variations in the testing data set probability. Excluding the BRAM usage, we can see that the \ee\ resource overhead is higher for design point A2, suggesting that the throughput is more tightly coupled to the performance of the first stage of the network.

\begin{table}
\centering
\caption{Comparison against BranchyNet reported accuracy, converted throughput and converted hard sample probability.}
\resizebox{0.4\textwidth}{!}{%
\begin{tabular}{rl|c|r|r}
\multicolumn{2}{c|}{\textbf{Network}} & \textbf{\begin{tabular}[c]{@{}c@{}}Top 1 \\Acc. (\%)\end{tabular}} & \textbf{\begin{tabular}[c]{@{}c@{}}$p$\\ (\%)\end{tabular}} & \textbf{\begin{tabular}[]{@{}c@{}}Throughput\\ (Sample/s)\end{tabular}} \\ \hline \hline
CPU & LeNet & 99.20 & - & 297 \\ 
CPU & B-LeNet & \textbf{99.25} & 5.7 & 1613 \\ \hline
GPU & LeNet & 99.20 & - & 633 \\ 
GPU & B-LeNet & \textbf{99.25} & 5.7 & 2941 \\ \hline
Baseline & LeNet & 98.84 & - & 43384 \\ 
\textbf{ATHEENA} & \textbf{B-LeNet} & 98.88 & 25.0 & \textbf{94170}
\end{tabular}%
}
\label{branchynet-cmpr}
\vspace*{-5mm}
\end{table}

The original BranchyNet paper details an implementation of the LeNet and Branchy (B) LeNet networks using using a 3.0GHz CPU with 20MB L3 Cache and NVIDIA GeForce GTX TITAN X (Maxwell) 12GB GPU. They report the average latency of a single sample in milliseconds. We convert their per sample average latency metric into a  throughput metric for the comparison in Table \ref{branchynet-cmpr}. Both our baseline and ATHEENA designs benefit from adaptations of the network architecture to be more amenable to a hardware implementation. These changes include quantisation to a fixed-point representation, adjusting layer parameters, and adapting the exit condition computation. This has a marginal effect on the accuracy compared to the software implementations and is partly the reason for such high gains compared the CPU and GPU implementations. We are able to exploit per-sample parallelism more effectively using the streaming architecture however the relative gains of implementing \ee\ on CPUs and GPUs is greater than demonstrated by our toolflow in part due to necessary reduction of exit percentage to maintain accuracy. Despite this, we find that a $p = 50\%$ still results in a throughput improvement relative to the baseline in spite of the area overhead for the additional layers and control logic embedded in the \ee\ streaming architecture. Overall, significant gains in throughput can be achieved from utilising an optimized streaming architecture converting a CNN implementation to FPGAs and up to $2.17\times$ further gains from implementing customised \ee\ hardware tailored to the design.

\subsection{Benchmarking Results}

We include an additional two networks in Table~\ref{benchmarking}. The best performing predicted throughput and corresponding resource results have been taken from the optimizer stage of the baseline and \owl\ toolflows. Due to the increased size of these networks we specify the target platform as the Xilinx VU440. We use the percentage of hard samples outlined in the papers to generate the two-stage design. We have included software-based implementations of these networks in the open source repository.

\begin{table}
\caption{Resulting throughput improvement for two-stage accelerator designs generated by \owl\ model compared to \fcn\ baseline. *Implemented on Xilinx ZC706}
\label{benchmarking}
\begin{tabular}{l|l|lc|c|r}
\multicolumn{1}{l|}{\textbf{Network}} & \textbf{\begin{tabular}{l}Toolflow\end{tabular}} & \multicolumn{2}{c|}{\textbf{\begin{tabular}[c]{@{}c@{}}Limiting\\ Resource\end{tabular}}} & \textbf{$p$} &  \begin{tabular}{c} \textbf{Throughput} \end{tabular} \\
\multicolumn{1}{l|}{\textit{(Task)}} &  & Type & \multicolumn{1}{l|}{\%} & (\%) & (Samples/s) \textit{Gain} \\ \hline \hline
B-LeNet\cite{branchynet} & Baseline* & \multicolumn{1}{l}{DSP} & 84 & - & 43384 \hskip 0.03\linewidth \textit{1.00$\times$} \\
\textit{(MNIST~\cite{deng2012mnist}} & ATHEENA* & \multicolumn{1}{l}{DSP} & 88 & 25 & \textbf{94170} \hskip 0.03\linewidth \textit{2.17$\times$} \\ \hline
Triple Wins\cite{triple_wins} & Baseline & \multicolumn{1}{l}{DSP} & 86 & - & 19524 \hskip 0.03\linewidth \textit{1.00$\times$} \\
\textit{(MNIST~\cite{deng2012mnist})} & ATHEENA & \multicolumn{1}{l}{DSP} & 81 & 25 & \textbf{54220} \hskip 0.03\linewidth \textit{2.78$\times$} \\ \hline
B-AlexNet\cite{branchynet} & Baseline & \multicolumn{1}{l}{DSP} & 84 & - & 8676 \hskip 0.03\linewidth \textit{1.00$\times$} \\
\textit{(CIFAR10~\cite{cifar10})} & ATHEENA & \multicolumn{1}{l}{DSP} & 88 & 34 & \textbf{17357} \hskip 0.03\linewidth \textit{2.00$\times$}
\end{tabular}
\vspace*{-4mm}
\end{table}

\section{Conclusions}
We have demonstrated the benefits of the \textit{input-dependent} computation paradigm in improving CNN mapping to FPGAs by developing a toolflow that allows for the exploration of the throughput-area trade-off space of \ee\ network hardware implementations orthogonal to benefits from quantisation and pruning employed by other frameworks. We have proposed an approach to automate the production of \ee\ networks based on a probabilistic proportioning of resources between parts of the computation operating at different data rates, and expanding an existing toolflow. This is achieved with development of specific \ee\ layers that can handle the intermediate buffering and conditional data-flow requirements of these networks.
We verify the toolflow's model of predicted performance by implementing multiple, resource-constrained, designs points on an FPGA board with randomised test samples. The robustness of the approach is explored using adapted test sets with known \ee\ probability variation. 
The resulting \owl\ framework can transform high-level \ee\ CNNs into optimized FPGA hardware that out perform their standard CNN counterparts in terms of throughput for a given board or area constraint.

\section*{Acknowledgment}
We would like to thank Alexander Montgomerie-Corcoran for the invaluable advice and assistance he has given with regard to the fpgaConvNet toolflow.

For the purpose of open access, the authors have applied a Creative Commons Attribution (CC BY) license to any Accepted Manuscript version arising.

\bibliographystyle{IEEEtran}
\bibliography{sample-base}

\end{document}